# Substrate Induced Molecular Conformations in Rubrene Thin Films: A Thickness Dependent Study

Sumona Sinha[‡], Chia-Hsin Wang[§], Manbendra Mukherjee*[‡] and Tapas Mukherjee[†]

[‡]Saha Institute of Nuclear Physics, 1/AF, Bidhannagar, Kolkata – 700064, India

[§]National Synchrotron Radiation Research Center, Hsinchu, Taiwan-30076

[†]Physics Department, Bhairav Ganguly College, Kolkata-700056, India

**ABSTRACT**

A systematic experimental and theoretical study about substrate induced molecular conformation in rubrene thin films by varying film thickness from sub-monolayer to multilayer, which currently attracts substantial attention with regard to its application in organic electronics, is performed. The clean polycrystalline Au and Ag were used as noble-metals, whereas, H passivated and $SiO_2$ terminated Si (100) were used as dielectric substrates. Angle dependent near edge x-ray absorption fine structure spectroscopy (NEXAFS) was employed to understand the molecular conformation whereas atomic force microscopy (AFM) was used to investigate the surface morphologies of the films. X-ray absorption spectra (XAS) of rubrene molecules with flat and twisted conformations were calculated using density functional theory (DFT). All the observed NEXAFS spectra of rubrene thin films at various thicknesses and onto different substrates were explained in terms of different combination of the spectral intensity from the twisted and the flat molecules. In contrast to general findings, comparatively rough polycrystalline Ag surface is found to support growth of better quality rubrene films. The results have important implications for the understanding of the substrate induced molecular conformations in rubrene thin films with its thickness and are beneficial for the improvement of the device performance.

- **INTRODUCTION**

Recently, organic electronic technologies are emerging as prospective future options for their high application potential in electronic devices. In contrast to inorganic semiconductors, whose physical principles of growth have been solidly established and successfully exploited to control the fabrication of desired morphologies and structures, the present-day knowledge on the growth of ordered organic films is still limited [1, 2]. Rubrene ($C_{42}H_{28}$) has been recently identified as a promising material due to its high charge-carrier mobility. Organic field effect transistors (OFET) based on rubrene single crystal currently hold the record of hole mobilities (~ 40 $cm^2$/V.s) among all organic semiconductors [3]. This remarkably high carrier mobility is ascribed to efficient cofacial $\pi$-$\pi$ stacking in the single crystal [4, 5]. Unfortunately, rubrene thin film transistors (TFTs) prepared by conventional organic molecular beam deposition techniques did not exhibit superior carrier mobilities, ranging from $10^{-6}$ to $10^{-2}$ $cm^2$/V.s, because of difficulty in achieving highly-quality crystalline rubrene active layer [6-8]. It is important to understand the growth process in detail to realize better-quality rubrene thin films. Rubrene is a nonplanar and flexible molecule comprising of a tetracene backbone core with two pairs of phenyl substituents symmetrically attached on either side of the backbone. The intra-molecular steric hindrance between the phenyl side groups results in a significant strain within the molecule which is minimized in two stable configurations, one with a twisted (figure 1 (a)) and the other with a flat tetracene backbone (figure 1 (b)) [9-11]. The twisting of the tetracene backbone is responsible for a complicated three dimensional structure of rubrene molecule which hampers the crystal formation. Moreover, our theoretical calculation showed that rubrene molecule with twisted backbone has lower total energy of about 285 meV compared to that with flat backbone[12]. In this context, it may be mentioned here near edge x-ray absorption fine structure spectroscopy (NEXAFS) is a very powerful tool for determining molecular conformation and orientation in organic films, even at submonolayer thicknesses [13]. Earlier, D. Kafer et al [9] and L. Wang et al [14] aimed to explain the role of molecular conformations in growth of rubrene thin films by using NEXAFS technique. The first group [9] observed the appearance of a peak, named $\alpha'$, at lower photon energy at lower thickness in addition to five main peaks of rubrene on Au (111) and various silicon substrates. This peak was found to disappear above a critical film thickness of about 12 nm. They assigned the $\alpha'$ peak to be originated from the twisting of backbone of rubrene molecules on the basis of a calculation with tetracene molecule [15]. The disappearance of $\alpha'$ peak with thickness was explained by the fact that the energy needed to planarize the backbone was compensated by lattice energy of molecular packing. On the other hand, Li Wang [14] et al could not find $\alpha'$ peak for rubrene films on Si (111). It should be noted that until now no theoretical reports on the calculation of NEXAFS spectra of different molecular conformation of rubrene molecules is available in the literature that can unambiguously explain the NEXAFS data of rubrene molecules obtained in different experiments.

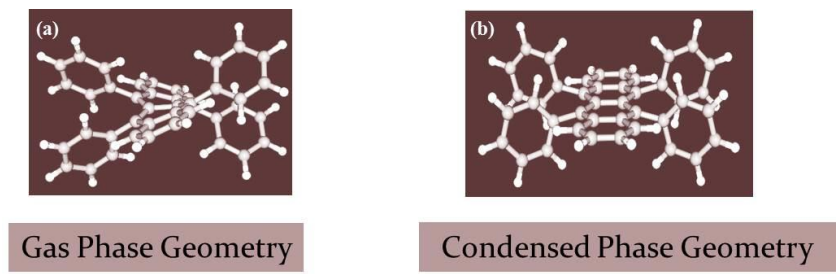

Figure 1: Rubrene molecule's (a) twisted conformation in gas phase and (b) flat conformation in condensed phase.

To resolve some of the apparent inconsistencies between the different reports on the growth of rubrene on weakly interacting substrates and to derive a more detailed understanding of the substrate induced molecular conformations in rubrene films with thickness we have carried out a comprehensive experimental and theoretical study on the molecular conformation in rubrene thin films by varying film thickness from sub-monolayer to multilayer on polycrystalline Au, polycrystalline Ag, $SiO_2$ terminated Si (100) and H passivated Si (100) substrates. We employed angle dependent near edge x-ray absorption fine structure spectroscopy (NEXAFS) and atomic force microscopy (AFM) for this purpose. X-ray absorption spectra (XAS) spectra of rubrene molecules with flat and twisted conformations were calculated using density functional theory (DFT). We have shown here that all the observed NEXAFS spectra of rubrene thin films at various thicknesses and on different substrates can be explained in terms of different combination of the spectral intensity from the twisted and the flat molecules. Furthermore, in contradiction -to the general observation, we have found that comparatively rough polycrystalline Ag surface supports growth of better quality rubrene films; similar to our previous observation with various cadmium arachidate (CdA) Langmuir −Blodgett (LB) multilayer covered $SiO_2$/Si (100) substrates [6]. Our results have significant inferences for the understanding of the substrate induced molecular conformations in rubrene thin films with thickness and are beneficial for the rubrene based device performance.

## ■ EXPERIMENTAL DETAILS

The NEXAFS experiments were carried at Beam-line (BL24A) of the storage ring of the National Synchrotron Radiation Research Center (NSRRC), Taiwan. The organic vapor deposition chamber was directly attached to the analysis chamber at the BL24A station allowing the samples to be studied without breaking the vacuum. We have used clean polycrystalline (poly) Au and Ag as noble-metal substrates. The gold coated Si substrate was purchased from Sigma Aldrich and the Ag coated (~75 nm) substrate was prepared by depositing Ag thin films on H terminated Si (100) using DC magnetron sputtering (PLS 500, Pfeiffer). H terminated Si (100), designated as Si-H, was prepared by etching the native oxide of Si (100) with 10% hydrofluoric acid for about 3 min. The $SiO_2$ coated (~100 nm) Si (100), designated as Si-O, was prepared by thermal oxidation of Si (100) and was cleaned by sonication in chloroform and methanol for 10 min each. The substrates were loaded into the UHV chamber immediately after preparation. To obtain clean metal substrates $Ar^+$ sputtering was used until the C 1s and O 2p XPS signals were vanished and consistent values of the work functions were obtained. Rubrene films were grown in a stepwise manner by thermal evaporation of rubrene powder (Acros Organics, 99%) from a resistively heated quartz crucible of a home-made organic material effusion cell. The nominal thickness of the rubrene films were calibrated by monitoring the evaporation rate with a quartz crystal microbalance (Inficon XTC controller). The evaporation rate for all the films was kept within 0.04-0.06 Å/s. For a systematic comparison, rubrene thin films with the same thickness but on different substrates (Au, Ag, Si-O and Si-H) were prepared together in the same chamber. After each deposition the samples were characterized by in-situ NEXAFS spectroscopy. To overcome the difficulty of performing a reliable C- K edge NEXAFS measurement for ultra-thin adsorbate films, the incident photon flux monitoring was done by measuring ion current of the ion chamber situated between the beam-line and the sample chamber, instead of traditional gold mesh method. The ion chamber is filled with Ar up to a working pressure of $10^{-3}$ torr and terminated with 0.1 µm thick Ti foils at both ends [16]. The NEXAFS spectra were measured in the partial electron yield (PEY) detection mode using a homemade electron detector based on a micro-channel plate and a retarding field of -150 V to optimize the signal to noise ratio. The polarization dependent NEXAFS spectra were obtained by varying the X-ray incident direction between normal incidence (90°) and glancing incidence (20°). The base pressure of deposition and analysis chambers were 1.0 x $10^{-8}$ torr and 1.0 x $10^{-10}$ torr respectively. The use of ion chamber for monitoring the incident photon flux also simplifies the NEXAFS data normalization schemes. The step of so-called mesh current normalization is not required here. Rest of the raw data background correction and normalization were carried out according to the established procedure [17].

Surface morphologies of the films were studied by using tapping mode AFM (Innova, Veeco) in air. The images were analyzed by using WSxM software [18]. Scans of different area over several regions of the films were taken to check the consistency of the morphology of the samples. All the depositions and characterizations were performed at room temperature.

## ■ COMPUTATIONAL METHODS

The equilibrium geometry and XAS spectra of the free molecule were calculated by DFT with the computer code STOBE-deMon [19, 20]. A gradient corrected RPBE exchange/correlation functional was applied [21, 22]. To calculate the equilibrium geometry and the X-ray absorption spectra, we used all-electron triple-ζ valence plus polarization (TZVP) atomic Gaussian basis sets for carbon centers, while the hydrogen basis sets were chosen to be of the double-ζ (DZVP) type [23]. The starting geometries for the optimization procedure were obtained using Avogrado (http://avogadro.cc/). To calculate X-ray absorption spectra, the Slater transition state method was applied [24, 25]. In this case the optimized geometry obtained from the geometry optimization calculation was kept fixed and absorption spectra were calculated. In order to obtain an improved representation of relaxation effects in the inner orbitals, the ionized center was described by using the IGLO-III basis [26]. A diffuse even tempered augmentation basis set was included at the excitation center to account for transitions to unbound resonances. The XAS spectra were generated through a Gaussian convolution of the discrete spectra with a broadening of 0.5 eV.

**Results and Discussions:**

Angle dependent C K-edge NEXAFS of 200 Å thick rubrene films on poly Au, poly Ag, Si-H and Si-O are shown in figure 2 (a) - (d) respectively. The spectrum generally consist of five well

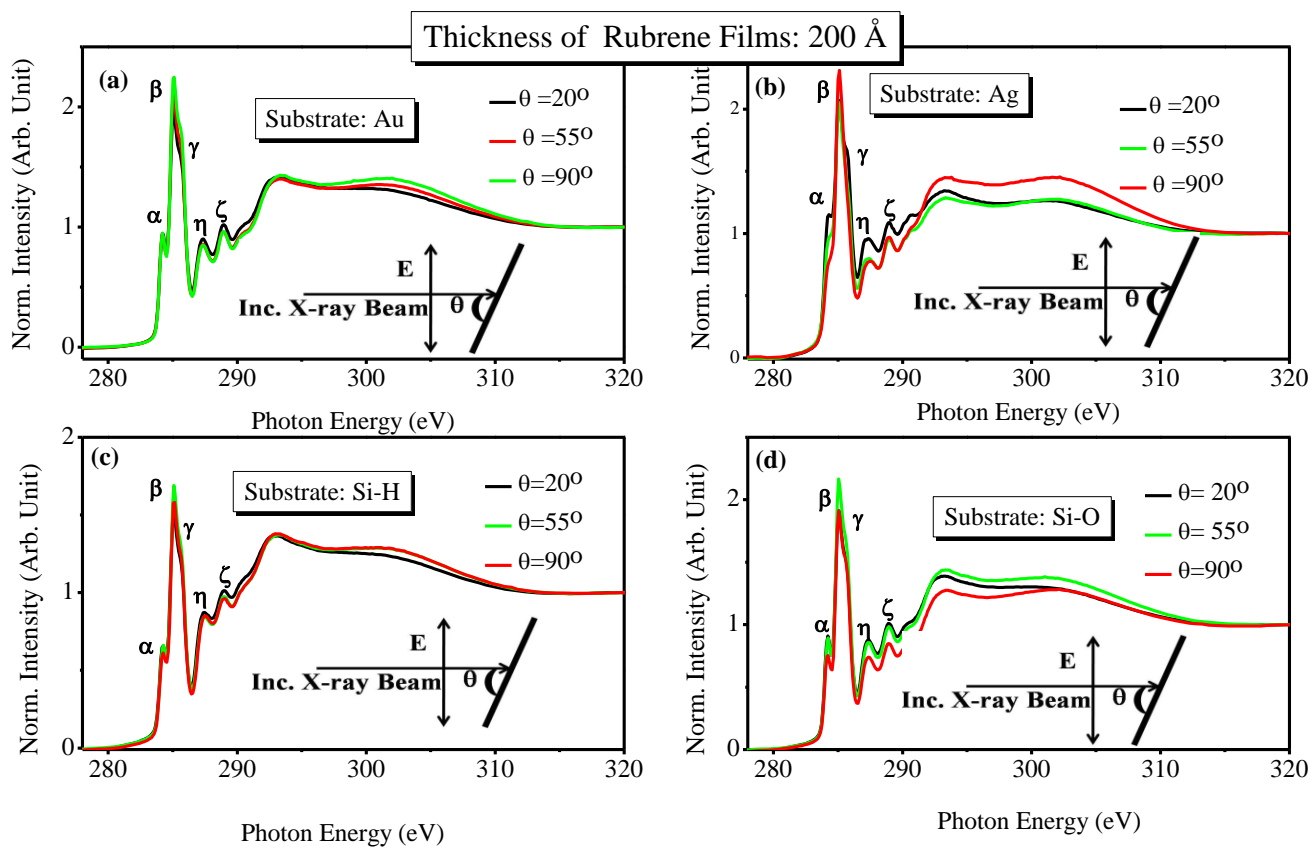

Figure 2: Angle dependent C K-edge NEXAFS of 200 Å thick rubrene layer on (a) poly Au, (b) poly Ag, (c) Si-H and (d) Si-O.

defined peaks, *α, β, γ, η* and *ζ* located at about energies of 284.3, 285.1, 285.6, 287.4 and 289.0 eV respectively. Earlier researchers have assigned [9, 14] the peaks labeled *α* and *γ* to the transitions from the carbon atoms at the tetracene backbone whereas the peak labeled *β* was associated to the excitations within the four phenyl side groups of rubrene. The two broad peaks labeled *η* and *ζ* were assigned to be from both the backbone and the side groups of the rubrene molecule. The peak positions are almost similar on four studied substrates and are close to the previously reported data [9, 14]. It can be observed from the figure that the change of normalized intensity of NEXAFS spectrum with incidence angle is not significant for all studied cases. The small but distinct linear dichroisms of the intensities of the *α* peaks for rubrene molecules on Ag substrates can be noticed. As both α and γ peaks are originated from the backbone and only β peak can be observed distinctly in the NEXAFS spectra, we plot the intensity of the *α* resonances of rubrene

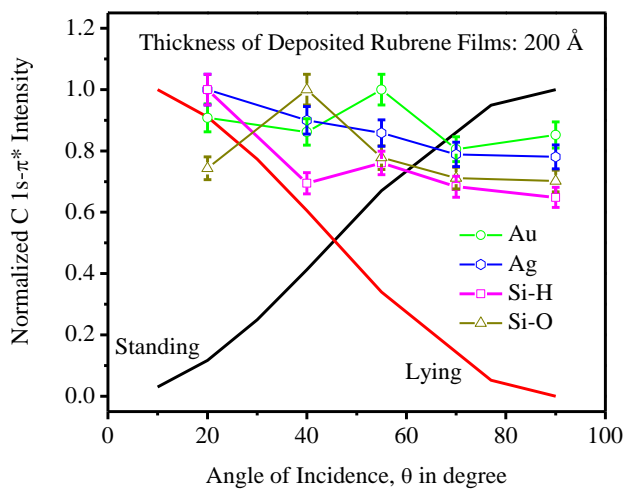

Figure 3: Angle dependence of normalized intensity of the $\alpha$ resonances of rubrene (film thickness: 200 Å) on (a) poly Au, (b) poly Ag, (c) Si-H and (d) Si-O. The dotted and dashed lines give the calculated angle dependence for ideally standing and lying molecules.

(film thickness: 200 Å) with angle of X-ray incidence [13] on poly Au, ploy Ag, Si-H and Si-O substrates in the figure 3. Small but systematic variation of intensity from carbon atoms at the tetracene backbone with incidence angle of radiation on poly Ag substrate suggests that tetracene backbone of rubrene molecules are in lying configuration on the Ag substrate. Whereas, change of peak intensity was not systematic on other substrates. This indicates that the molecules are randomly oriented on poly Au, Si-H and Si-O substrate. In this context, it can be mentioned here that we have performed X-ray diffraction (XRD) measurements on four samples and observed that the films are amorphous in nature (data not shown here). It can be explained by the fact that nanocrystalline films are formed for our kind of preparation as has been previously pointed out by others [9, 14]. Moreover, such nano dimentional crystallites are rather immobile at room temperature and does not aggregate further to form large crystals.

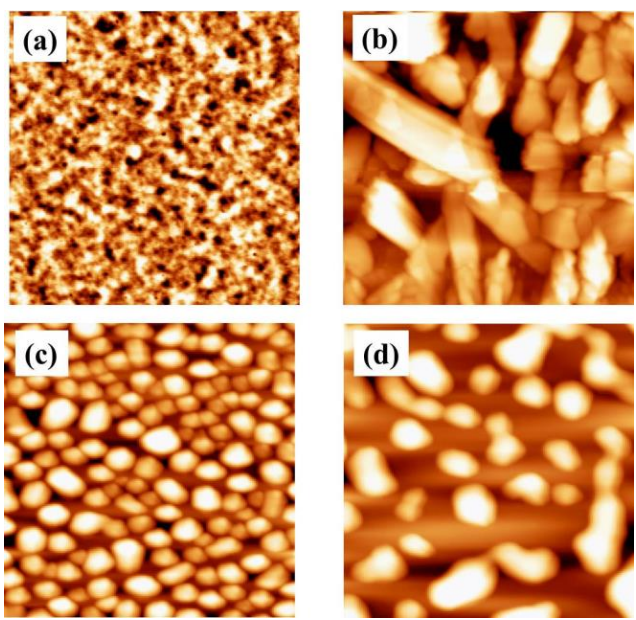

Figure 4: AFM topography images (size: 5 µm x 5µm) of 200 Å thick rubrene films on (a) poly Au, (b) poly Ag, (c) Si-H and (d) Si-O substrates.

Under this circumstance, we have decided to check the surface morphology of these deposited rubrene films on various substrates; the AFM imaging experiment was carried out. Figure 4 displayed the AFM images of 200 Å thick rubrene layers on four different substrates respectively. Some differences in surface morphology of rubrene films grown on four different substrates can be easily observed from the figures. The films on poly Au is found as flat and grainy whereas a film with comparatively larger grains was observed on Si-H. On Si-O substrate the grain density was lower compared to all other cases. A better rubrene film with larger closely packed nanocrystalline (upto 1 or 2 micron) grains was noticed on poly Ag substrate. This observation along with NEXAFS studies (figure 1 and 2) indicates that the order of nanocrystallinty is higher for 200 Å thick rubrene film on poly Ag, compared to other three cases.

Moreover, we have shown the angle dependent NEXAFS spectra for 50 Å and 20 Å thick rubrene layer on four different substrates (shown in figure S1 and S2 respectively, Supporting

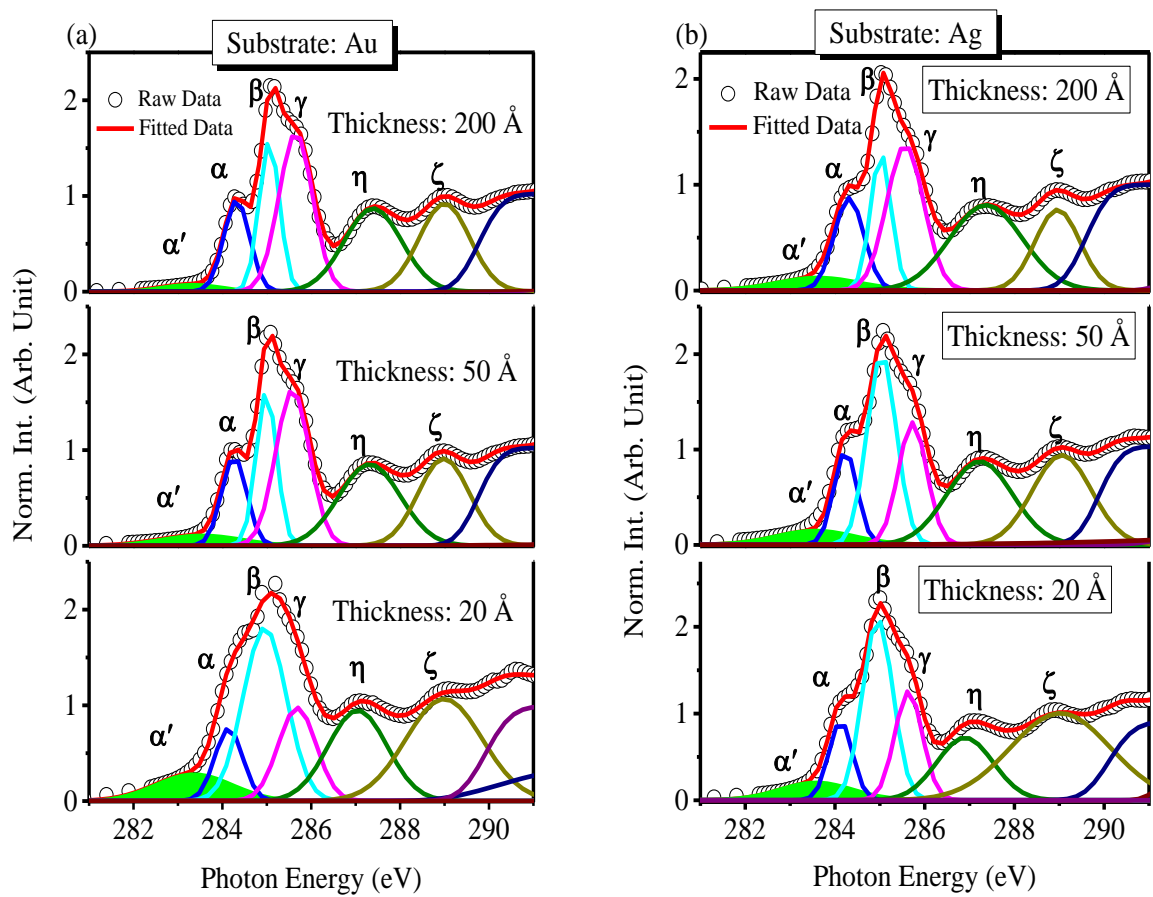

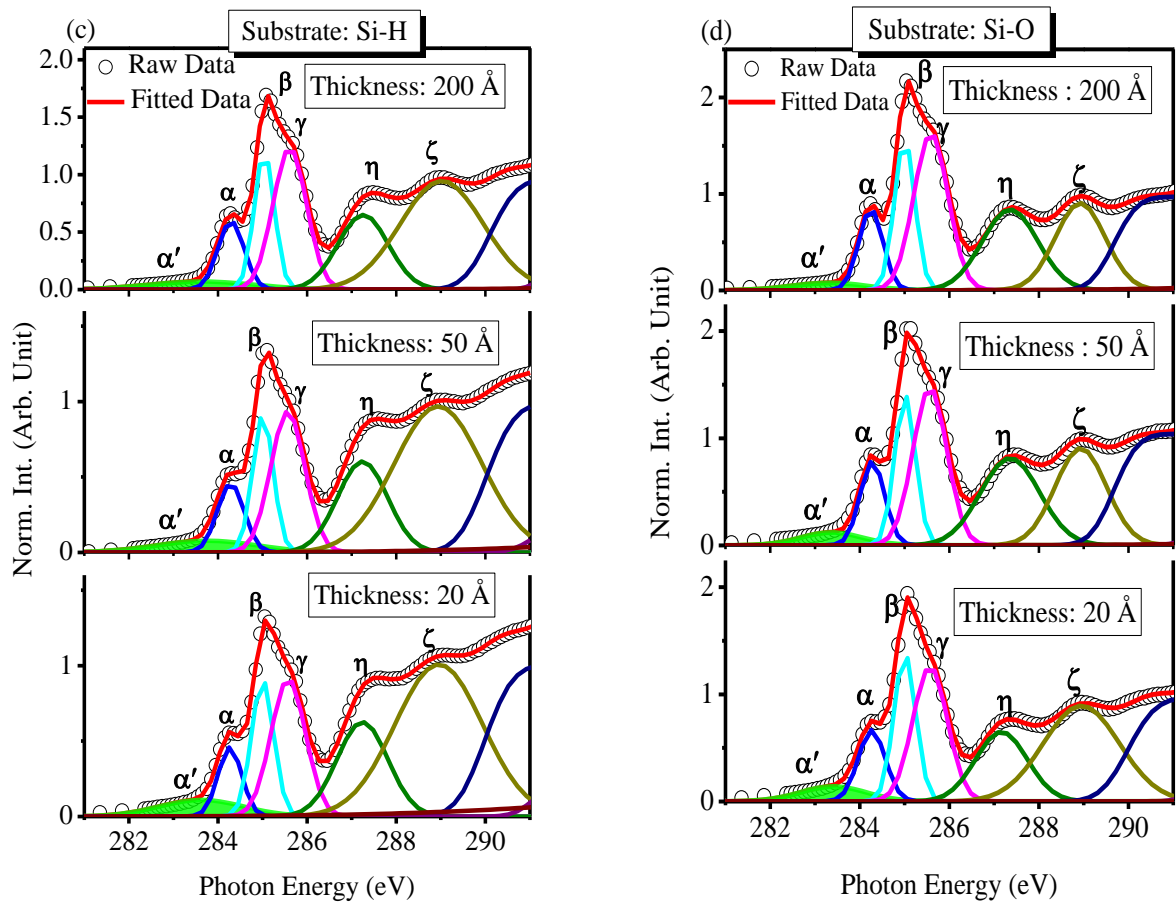

Figure 5: Fitted $\pi^*$ region of C K-edge NEXAFS spectra of 55° incidence angle of 20 Å, 50 Å and 200 Å thick rubrene layers on (a) poly Au, (b) poly Ag, (c) Si-H and (d) Si-O substrates.

Information). No prominent linear dichroism was observed in all the studied cases. These suggest that rubrene molecules are randomly oriented at lower thickness on these four weakly interacting substrates. But a noticeable change in NEXAFS spectra with thickness of rubrene layer is observed. To study the evolution of the shape of NEXAFS spectra with thickness on four different substrates, the fitted $\pi^*$ region of NEXAFS spectra taken at an incidence angle of 55° for 20, 50 and 200 Å thick rubrene layer on four substrates are demonstrated in figure 5 (a)- (d) respectively. In order to identify changes to the absorption features, all spectra are fitted using the same procedure with Gaussian peaks (corresponding to the $\pi^*$ excitations) and asymmetric Gaussian peaks (representing the $\sigma^*$ transitions) and a step function for the background [13]. In addition to the five prominent resonances, an extra peak (labeled $\alpha'$) at an energy of ∼ 283.5 eV is needed to fit the spectra. To compare systematically, the evolution of the $\alpha'$ peak intensity (area under the fitted curve) with increasing film thickness on four different substrates is plotted in figure 6 (a). The intensity of $\alpha'$ peak is reduced with increasing molecular layer thickness on all substrates. The reduction of intensity of $\alpha'$ peak is more prominent for rubrene on poly Au, the reduction was about 79.0 % with increase of layer thickness from 20 to 200 Å. This numbers are 38.5%, 27.5% and 19.2% on Si-O, poly Ag and Si-H substrate. As the $\alpha$ peak can be observed distinctly in the NEXAFS spectra, we examine carefully the evolution of $\alpha$ peak intensity with thickness of rubrene layer on four different substrates and the result is shown in figure 6 (b). The intensity of $\alpha$ peak is enhanced with thickness on four different substrates. Simultaneous reduction of $\alpha'$ peak intensity and enhancement of $\alpha$ peak intensity indicates that these two opposite phenomena may be correlated. In order to substantiate theses experimental findings and to understand the situation in a better way, we have carried out theoretical investigations. The calculated C K-edge angle integrated XAS spectra for 100% flat (black line), 85% flat + 15% twisted (blue line), 50% flat + 50% twisted (brown line) and 30% flat + 70% twisted (olive line), 100% twisted (red line) weighted random rubrene molecules are shown in figure 7. It can be observed for 100% twisted rubrene molecules that distinct peaks are located at 283.5 eV, 285.3 eV, 286.8 eV and 289.0 eV from the figure. The peaks can be assigned by comparing our experimental NEXAFS spectrum as α′, β, η and ζ peaks. Mainly the intensity of α′ and β was gradually reduced as well as two new distinct peaks was progressively appeared at 284.4 eV and 285.8 eV with reduction of twisted conformation weight. These two peaks can be ascribed as α and γ peaks. The α′ peak was found to completely disappear for 100% flat rubrene molecules. Moreover, a negligible shift towards lower photon energy was observed with reduction of twisted conformation weight. Therefore our theoretical study indicates that (i) the α′ peak appears for excitations within the twisted backbone and (ii) more prominent α peak is the signature of flat tetracene backbone of the rubrene molecule. The reduction of $α′$ peak intensity and the enhancement of $α$ peak intensity (figure 6) suggest that higher percentage of molecules attain flat backbone molecular conformation with increase in film thickness. We already mentioned that rubrene molecule with twisted backbone has lower total energy compared to that with flat backbone. The energy difference obtained from calculation was about 285 meV [12]. This energy is needed to planarize the backbone. Here this energy compensation may occur from substrate surface interaction and molecular packing. Moreover, the additional peak (α') for the twisted rubrene molecule is likely to originate due to the proximity effects of the atoms in some part of the molecules as a result of twisting. It was also observed from our DFT calculation, that α' peak appears from the two central carbon atoms of the twisted rubrene molecule. The transition responsible for the peak was assigned to be from 1s to LUMO+1 level of the molecule.

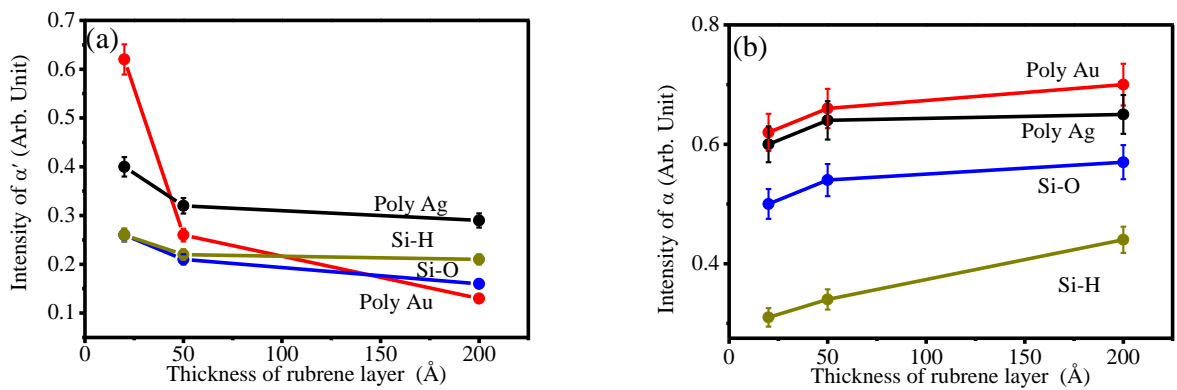

Figure 6: Intensities of (a) $\alpha'$ and (b) $\alpha$ peak of C K-edge NEXAFS spectra of 55° incidence angle of rubrene layers on poly Au, poly Ag, Si-H and Si-O substrates.

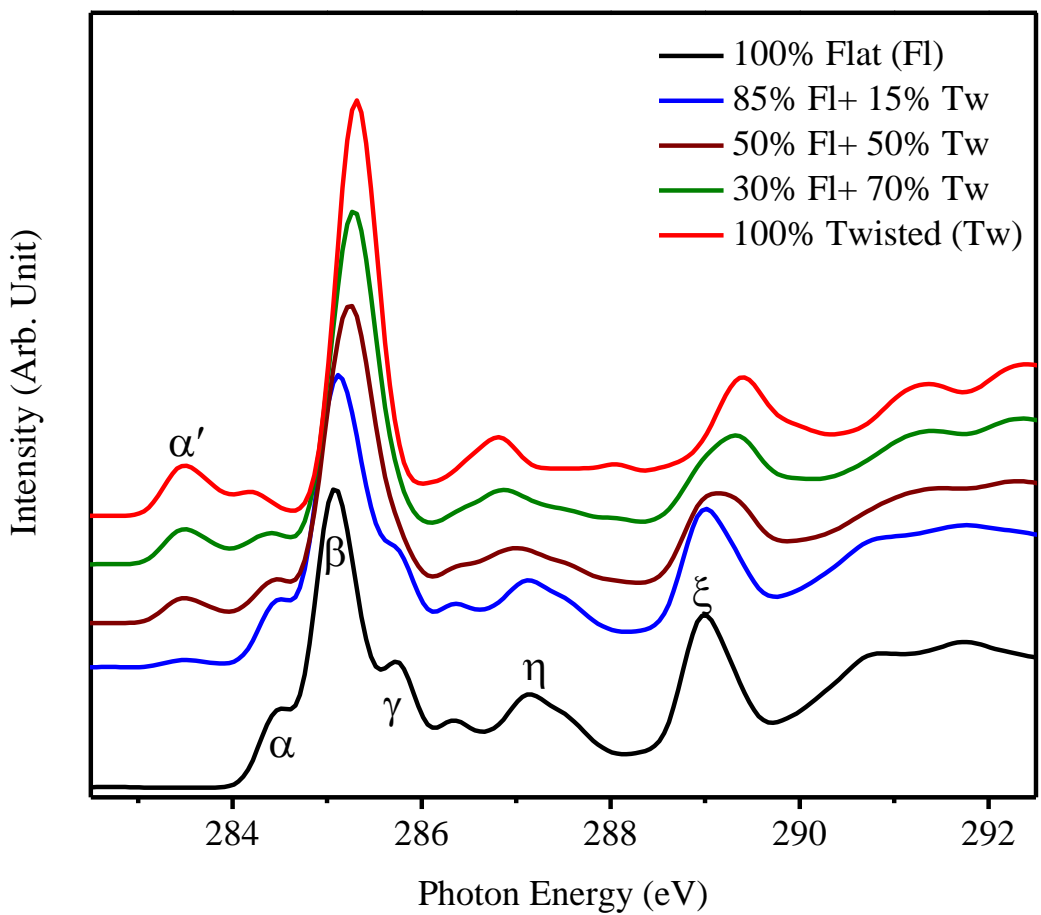

Figure 7. Calculated C K-edge angle integrated XAS spectra for 100% flat (black line), 85% flat + 15% twisted (blue line), 50% flat + 50% twisted (brown line) and 30% flat + 70% twisted (olive line), 100% twisted (red line) weighted random rubrene molecules. The vertical scales are adjusted to better show the spectral features.

Another attention-grabbing feature was observed with NEXAFS spectra for sub-monolayer thick rubrene films. The NEXAFS spectra for sub-monolayer thickness of rubrene molecules on four different substrates show some interesting features. Figure 8 (a) displays C K-edge NEXAFS spectra at 55° incidence angle of 5 Å thick rubrene layers on poly Au, poly Ag, Si-H and Si-O respectively. The relative intensities of the peaks at $\pi^*$ region of the spectra dramatically vary at lower thickness compared to those with higher thickness for different substrates. This observation clearly shows that the interactions between the rubrene molecules and the substrates play crucial roles in the ordering and the molecular conformations. It can be observed from the figure that α and γ peaks (positions shown by the dotted lines) of 5 Å thick rubrene layers are less prominent compared to that of thick samples. Earlier we observed [12] NEXAFS spectra of similar shape from oxygen-reacted rubrene molecules. The oxygen atoms are attached to the backbone of rubrene and as a result the backbone becomes more twisted. Consequently the intensity of α and γ peaks (shown by dashed lines in the figure) were drastically reduced. It may be anticipated that the twisting of tetracene backbone at sub-monolayer thickness produces similar effect to some extent.

In order to validate the experimental findings and to understand the situation in a better way, we have carried out theoretical investigations. We have considered free randomly oriented rubrene molecule in flat and twisted conformation for theoretical calculation. As angle dependent

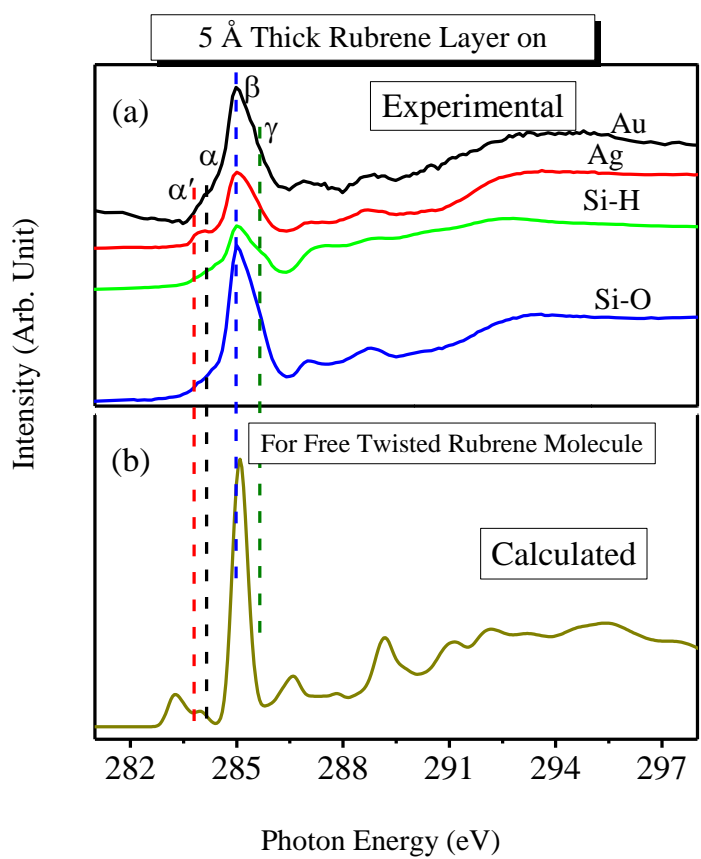

Figure 8: (a) C K-edge NEXAFS spectra of 55° incidence angle of 5 Å thick rubrene layers on poly Au, poly Ag, Si-H and Si-O substrates; (b) calculated angle integrated C K-edge XAS spectra for free twisted rubrene molecule.

C K-edge NEXAFS spectra of 5 Å thick rubrene layer on all studied substrate does not give any signature of linear dichroism (data not shown here), we have calculated angle integrated XAS spectra. The calculated angle integrated C K-edge XAS spectra for free twisted rubrene molecule is displayed in figure 8 (b). The calculated spectrum shows a fairly narrow single peak (β) with negligible intensity of α and γ peaks. Due to physical adsorption on these substrates, the deposited molecules are likely to retain the twisted geometry of the free rubrene molecule. The appearance of distinct α peak was noted on poly Ag substrate whereas this peak was not prominently distinguishable on other studied substrates. Furthermore, it can be mentioned here that we have observed systematic molecular orientation (figure 2 and figure 3) along with closely packed crystal-like features (figure 4) for 200 Å thick rubrene layer on poly Ag substrate. These results suggest that quality of rubrene film was better on poly Ag substrate than on other three substrates in terms of compactness of larger and regular shaped grains. We have noted earlier that Poly Ag shows relatively larger reactivity compared to other noble metal substrates [27, 28]. Moreover, we had previously reported [27] that the rubrene films were grown on these four different substrates by following layer-plus-island growth mode similar to that of the majority of vapour deposited organic semiconductors on the weakly interacting substrates [29-31] from analyzing the evolution of X-ray photoemission spectroscopy (XPS) peak intensity with the deposited film thickness along with AFM images at lower thickness. The underlying smooth layer thickness below the islands was found to be about 10 Å. Similar value as a thickness of underlying layer for various vapour deposited organic semiconducting films on the weakly interacting substrates was earlier observed by several authors [29-31]. Combined effect of higher reactivity of Poly Ag along with higher roughness in this substrate may explain the appearance of relatively stronger a peak on Poly Ag substrate in figure 8 (a).

It is well-known that quality of vacuum deposited organic films is better on comparatively smooth substrate surfaces [32-34]. However, in our earlier study we found contradictory observation for rubrene films on various cadmium arachidate (CdA) Langmuir −Blodgett (LB) multilayer covered $SiO_2$/Si (100) substrates [6]. The better rubrene films were found to grow on relatively rough CdA LB layer coated $SiO_2$/Si surface. Motivated by our earlier results here we have studied the AFM images of the bare substrates for measuring the substrate roughness and to study if there was any correlation between the substrate roughness and the film quality. Figure 9 shows AFM images of various substrates that are used to support the rubrene growth. It can be observed from the images that poly Au, Si-H and Si-O substrate possess grainy features with random height variations. The line profiles show that the surface roughness was very low for Si-H and Si-O substrates. The values are in the range of 0-6 Å for Si-H and Si-O substrates, whereas, the range is found as 0-30 Å for deposited Au films. On the other hand, a completely different morphology was noted in case of poly Ag surfaces (figure 9 (b)). The image shows wide, flatter terrains separated by a large height difference of about 100 Å. It is interesting to note that the quality (in terms of crystallinity) of the rubrene film grown on the Ag substrate was the best among all the films as shown in figure 4 (b). These indicate that contrary to the general understanding, a surface comprised of wide, flatter terrains separated by large height differences may help to achieve better quality of film growth compared to a surface of lower roughness with grainy features and random height variations.

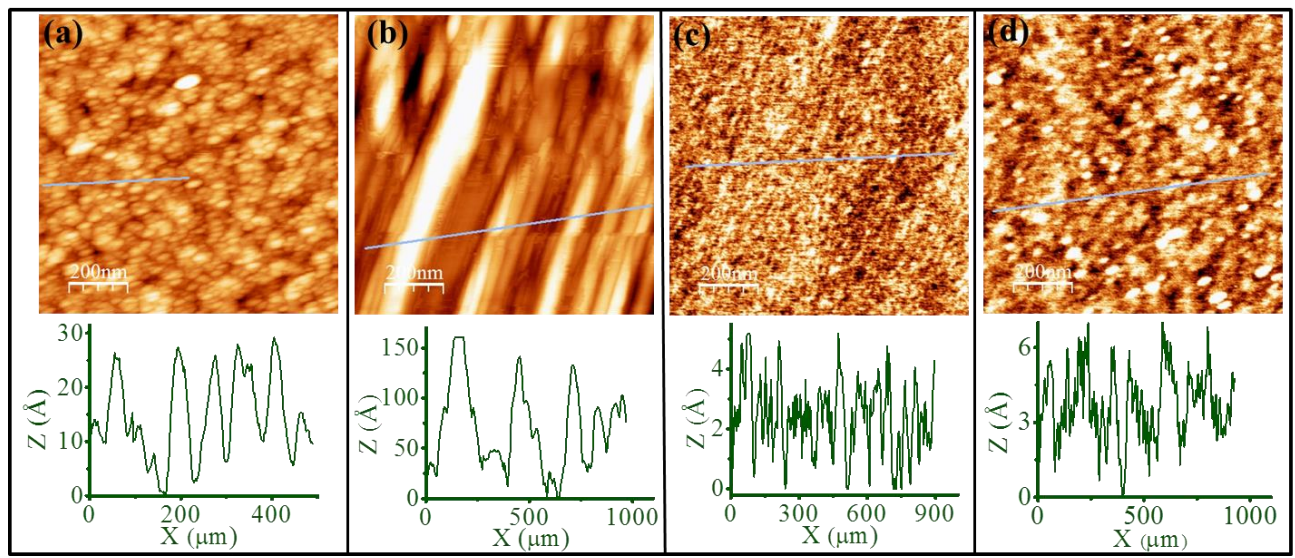

Figure 9. AFM images of various substrates used to support the rubrene growth: (a) poly Au, (b) poly Ag, (c) Si-H and (d) Si-O. The size of all of the images is 1 μm × 1 μm. Also shown beneath each AFM image is the height variation along the designated line in the image.

■ **CONCLUSION**

In conclusion, we have systematically studied the substrate induced molecular conformations in rubrene thin films by varying film thickness from sub-monolayer to multilayer on four different substrates; which presently attracts substantial attention with regard to its application in organic electronics. With the help of our theoretical calculation we explain all the observed NEXAFS spectra of rubrene thin films at different thicknesses and on various substrates by different combinations of the spectral intensity from the twisted and the flat molecules. Contrary to the general observations, we noted that comparatively rough polycrystalline Ag surface supports to grow better quality of deposited rubrene films. The results thus have significances for understanding of the substrate induced molecular conformations in rubrene films and are beneficial for the device performance.

■ **ASSOCIATED CONTENT**

*S **Supporting Information**

1) Angle dependent C K-edge NEXAFS of 50 Å thick rubrene layer on (a) poly Au, (b) poly Ag, (c) Si-H and (d) Si-O.

2) Angle dependent C K-edge NEXAFS of 20 Å thick rubrene layer on (a) poly Au, (b) poly Ag, (c) Si-H and (d) Si-O.

■ **AUTHOR INFORMATION**

**Corresponding Author**

*E-mail: manabendra.mukherjee@saha.ac.in.

**Notes**

The authors declare no competing financial interest.

■ **ACKNOWLEDGEMENTS**

The work is partially supported by the India-Taiwan program in Science and Technology.

■ **References**